\documentclass[ pra,a4paper,aps,twocolumn,superscriptaddress]{revtex4-1}
\usepackage{graphicx}
\usepackage{bm}
\usepackage[english]{babel}
\usepackage{amsmath,times}
\usepackage{amsfonts}
\usepackage{ulem}
\usepackage{soul}
\usepackage[pdftex,colorlinks=true,allcolors=blue]{hyperref}

\def\bra#1{\mathinner{\langle{#1}|}}
\def\ket#1{\mathinner{|{#1}\rangle}}
\def\tc#1{{\color{black}#1}}

\newcommand{\ketbra}[2]{\ket{#1}\bra{#2}}

\begin{document}
\title{Quantum probes for the cutoff frequency of Ohmic environments}
\author{Claudia Benedetti}
\affiliation{Quantum Technology Lab, Physics Department, Universit\`a degli Studi di Milano,  Milano, Italy}
\author{Fahimeh Salari Sehdaran}
\affiliation{Faculty of Physics, Shahid Bahonar University of Kerman, Kerman, Iran}
\author{Mohammad H. Zandi}
\affiliation{Faculty of Physics, Shahid Bahonar University of Kerman, Kerman, Iran}
\author{Matteo G. A. Paris}
\affiliation{Quantum Technology Lab, Physics Department, Universit\`a degli Studi di Milano, Milano, Italy}
\date{\today}
\begin{abstract}
Quantum probing consists of suitably exploiting a simple, 
small, and controllable quantum system to characterize a larger 
and more complex system. Here, we address the estimation of the 
cutoff frequency of the Ohmic spectral density of a harmonic 
reservoir by quantum probes.  To this aim, we address the use of 
single-qubit and two-qubit systems and different kinds of coupling with 
the bath of oscillators. We assess the estimation precision by 
the quantum Fisher information {\it of the sole quantum probe} as well
as the corresponding quantum signal-to-noise ratio. We prove that, \tc{for 
most of the values of the Ohmicity parameter}, a 
simple probe such as a single qubit is \tc{already optimal} for the precise 
estimation of the cutoff frequency. 
Indeed \tc{for those values}, upon considering a two-qubit probe either 
\tc{in a Bell} or in separable state, we do not find improvement to the 
estimation precision. \tc{However, we also showed that there exist few 
conditions where employing two qubits in a Bell state interacting with a common bath 
is more suitable for precisely estimating the cutoff frequency.}
\end{abstract}
\maketitle
\section{Introduction}
Complex quantum systems with many degrees of freedom are often difficult 
to access and, in turn, to characterize. A possible strategy to overcome 
this difficulty is that of monitoring only a small portion of the system 
and exploiting an indirect measurement scheme to estimate the value of the 
parameters of interest. 
An effective way to implement this paradigm is by means of quantum probes.
A quantum probe is a simple and controllable quantum system that interacts 
with a larger reservoir (also refereed to as an environment or bath) and becomes 
entangled  with it. Due to quantum correlations the probe becomes extremely
sensitive to the perturbations induced by the environment, and upon performing 
a measurement on the quantum probes one may effectively infer the properties 
of the environment 
\cite{breuer2002theory,abdelrahman,elliott,streif16,troiani16,cosco17}, i.e., extract 
information on the parameter of interest. The outcomes of the measurement 
performed on the probe are then used to build an estimator for the unknown 
parameter, whose precision can be assessed using  the tools of quantum 
estimation theory (QET) \cite{parisQE}. Indeed, QET has already proven useful
in different contexts, ranging from the estimation of the spectral properties 
of the environment \cite{benedetti2014,benedetti14, zwick16} to quantum channel 
parameters \cite{monras07,fujiwara01,fujiwara03,pinel13}, quantum correlations 
\cite{brida10, brida11, blandino12, benedetti13}, optical phase \cite{monras06, 
allevi, kacprowicz10, genoni11, spagnolo12}, quantum thermometry \cite{brunelli11,
correa15}, and the coupling constants of different kinds of 
interactions \cite{stenberg14,bina16,tuffa,tama16,nokkala17}. In particular, 
the quantum Fisher information (QFI) is the quantity that allows us to evaluate  
the ultimate precision of any estimation procedure as ruled by quantum mechanics through 
the quantum Cram\'er-Rao bound (CRB). The larger the QFI, the more accurate is the estimation 
strategy. 
\par
\tc{A relevant quantity to characterize complex environments is the so-called 
spectral density, which is the Fourier transform of its autocorrelation function 
and, in turn, determines how and how fast quantum probes are going to decohere.
In devices of interest for quantum technology this determines the available 
coherence time for communication and computation, and thus a precise characterization
of the spectral density is a crucial step to design engineered reservoirs.  
Thermal noise shows a flat spectrum, while in structured reservoirs as those
encountered working with Josephson junctions \cite{ast06}, or photonic crystals \cite{jan95}, 
different spectra may be observed. In this framework, a crucial parameter 
characterizing a complex environment is its cutoff frequency, 
which is linked to the environment correlation time as $\tau_c=1/\omega_c$, 
and represents the frequency above which the spectral density starts 
to fall off. }
\par
In particular, in this work we consider an exponential cutoff function 
and address the estimation of the cutoff frequency for the Ohmic family 
of spectral densities characterizing a bosonic reservoir. In order to pursue 
this task, we consider single- and two-qubit systems, interacting
with their environment and use them as quantum probes. This means optimizing 
the initial preparation of the probe and performing a measurement on the 
system to extract information about the spectral cutoff frequency. Due to 
the interaction with the environment, the quantum probes will be generally 
subjected to decoherence (dephasing) and dissipation phenomena. The   
timescales of these processes  depend on the physical context considered. 
Usually, the dissipation timescale is much longer than the decoherence one, 
such that the dynamics of many systems of interest may be described as pure 
dephasing \cite{benedettiDeph,addis2014}, and this is indeed the case 
considered here. 
\par
We  compare the behavior of the quantum Fisher information and the 
signal-to-noise ratio (SNR) for different values of the Ohmic parameter, 
moving from sub-Ohmic to super-Ohmic regimes. We first study the case of 
a single qubit used as a probe, then we extend our analysis to the two-qubit 
scenario, in both independent and common environments. In this way, we try 
to understand whether multiple (and entangled) probes may improve the estimation 
procedure.  We compare the maximized QFI at the optimal interaction time and 
prove that, for most of the Ohmicity parameter values, a single quantum probe is already 
sufficient to achieve optimal estimation of the parameter.
\par
The paper is organized as follows: In Sec. \ref{sec:2}, we introduce 
the physical model, whereas in Sec. \ref{sec:3} we briefly summarize 
the tools of local estimation theory. In Sec. \ref{sec:4}, we present 
our results on the precision achievable by quantum probes in the estimation 
of the cutoff frequency of the spectral density. Section \ref{sec:5} closes 
the paper with some concluding remarks.
%
\section{The physical model}\label{sec:2}
We consider a pure dephasing model consisting of one or two qubits which interact 
with a bosonic reservoir at zero temperature, characterized by an Ohmic spectral density. 
This model allows for an exact analytic solution \cite{breuer2002theory, palma96} and  
many of its features have already been analyzed  \cite{addis14, reina02,goan10, ban17, lampo17}. 
Here we change the point of view with respect previous studies, i.e., we use the qubits as  
quantum probes for a spectral parameter of the system-reservoir couplings, rather than looking
for the decoherence effects on the qubits assuming the knowledge of the reservoir.
\subsection{Single qubit}
We first focus on a single-qubit probe, characterized by energy spacing 
$\omega_0$, coupled with all the modes of a bath of harmonic oscillators 
(hereafter we set $\hbar=1$ \tc{and we scale all frequencies with $\omega_0$}).
The global \tc{dimensionless 
Hamiltonian $\mathcal{H}=\mathcal{H}_S+\mathcal{H}_B+\mathcal{H}_I$ is given by
\begin{equation}
\mathcal{H}=\frac{1}{2}\, \sigma_{z}+\sum_{k} \omega_{k}\,b_{k}^\dagger\,b_{k}+
\sum_{k}\sigma_{z}(g_{k}\,b_{k}^\dagger+g_{k}^\ast\,b_{k}),
\end{equation}
where $\sigma_z$ is the Pauli operator of the qubit, $b_k^{\dagger}(b_k)$ 
denotes bosonic creation (annihilation) operator for mode $k$, 
satisfying the commutation relation $ [b_{k},b_{k'}^\dagger]=\delta_{kk'}$,  
$\omega_k$ is the frequency of the $k$-th mode, and 
$g_k$ is the corresponding coupling constant with the qubit. Both, $\omega_k$ and $g_k$ are expressed in units of $\omega_0$ and are thus dimensionless.}
\par
The couplings $g_{k}$ can be distributed according to different spectral distributions, 
which lead to different dynamical properties for the qubit. 
Following \cite{breuer2002theory,palma96}, we can calculate the reduced 
dynamics of the qubit in the interaction picture.
We suppose that the bath is initially in a thermal state at zero 
temperature. If we move to a continuum limit  $\omega_k\rightarrow 
\omega(k)$ and $\sum_k\rightarrow \int d\omega f(\omega)$, with 
$f(\omega)$  the density of modes, we can 
introduce the  spectral density $J(\omega)=4 f(\omega) |g(\omega)|^2$.
Assuming that the couplings $g(\omega)$ are nearly constant in $\omega$,
$J(\omega)$ becomes the spectral density of the bath's modes.  
Here we consider a reservoir  with a spectral density 
belonging to the Ohmic class
 :
\begin{equation}
J(\omega,\omega_c)=\frac{\omega^{s}}{\omega_{c}^{s-1}}\,e^{-\frac{	\omega}{	\omega_{c}}},
\label{Jomega}
\end{equation}
parametrized by a real positive number  $s$, which moves the spectrum from sub-Ohmic
 ($s < 1$) to Ohmic ($ s = 1$) and super-Ohmic ($s > 1$) regime.  
 \tc{Common values of $s$ are $0.5,\,1,\,3$, used to describe  quantum Brownian motion,
 conductive electrons in metals, phonon baths, $1/f^{\alpha}$ noise in solids and in superconducting qubits,  and the interaction between a charged particle and its own electromagnetic field \cite{weiss, paavola, barone91, benedettif}.
 }
 $\omega_{c}$ is the cutoff frequency, i.e. the parameter we want to 
 estimate using quantum probes.
 Once the spectral density is fixed, the qubit dynamics can be easily  
 calculated through the
 single qubit quantum map $\Phi(t)$:
 \begin{equation}
\rho(t)=\Phi(t)\circ\rho(0),
\end{equation}
where
\begin{equation}\Phi(t) =\left( \begin{matrix}
1 & e^{-\Gamma(t,\omega_{c})} \\ e^{-\Gamma(t,\omega_{c})} & 1 \end{matrix}\right),
\label{phi1}
\end{equation}
where $\rho(0)$ is the initial state of the qubit, \tc{$t$ is the dimensionless time } and $\circ$ is  the element-wise Hadamard product  \cite{addis14}. 
The decoherence factor $\Gamma(t,\omega_c)$ depends upon the spectral density of the bath and takes the form:
\begin{equation}
\Gamma(t,\omega_{c})=\int_0^{\infty}  \frac{1-\cos(\omega t)}{\omega^2}\,J(\omega,\omega_c)\, d\omega.
\label{gamma1}
\end{equation}
The explicit expression of Eq. \eqref{gamma1}  depends  on the Ohmicity parameter $s$:
\begin{align}
\Gamma(t,\omega_{\!c} )& \!=\! \left\{\begin{array}{ll}
\frac12 \log\left(1+(\omega_{c} t)^2\right)&s=1\\
\\
\!\!\!\! \left(\! 1-\frac{\cos[(s-1)\arctan(\omega_{c} t)]}{\left[1+(\omega_{c} t)^2\right]^{\frac{s-1}{2}}} \! \right)\! \bar{\Gamma}[s-1]&s\neq1
\end{array}\right.
\label{gamtau}
\end{align}
where   $\bar{\Gamma}[x]=\int_0^\infty t^{x-1}e^{-t } dt$.
\subsection{Two qubits}
We are now going to analyze the case of two non-interacting qubits coupled with the bosonic reservoir. Two different scenarios  arise: either the two qubits are coupled to two independent local reservoirs, or they are embedded in the same bath.
\subsubsection{Two qubits in independent environments}
In the case of two non-interacting qubits coupled to independent but identical environments, the global Hamiltonian is:
 \begin{align}
 \mathcal{H}=\mathcal{H}^{(1)}+\mathcal{H}^{(2)}
 \end{align}
 where the \tc{dimensionless }single qubit Hamiltonian $\mathcal{H}^{(j)}$, $j=1,2$, is given by 
 \begin{align}
\!\!\! \mathcal{H}^{(j)}\! = \! \frac12 \sigma_z^{(j)} \!+ \! \sum_k \! \omega_k b_k^{\!\dagger (j)}b_k^{\!(j)} \!\! +\! \sum_k \! \sigma_z^{(j)} \!\! \left(g_k b_k^{\dagger (j)} \!\! + \!\! g_k^{*}b_k^{(j)}\! \right)
\end{align}
and we assume that the qubits are coupled to their respective baths with the same strengths $g_k^{(1)}=g_k^{(2)}\, \forall k$. 
The two-qubit density matrix has the form
\begin{equation}
\rho_I(t)=\Phi_I(t)\circ\rho(0)
\end{equation}
where the two-qubit map is the tensor product of the single qubit channel \eqref{phi1}:
\begin{equation}
\Phi_I(t)=\Phi(t)\otimes\Phi(t)
\end{equation}
and $\rho(0)$ is  the initial state of the two qubits.
\subsubsection{Two qubits in a common environment}
 We now assume that the the two qubits are now coupled to the same reservoir. The total Hamiltonian \tc{is}:
\begin{equation}
\mathcal{H}\!= \! \frac12 \! \sum_{j=1}^2 \! \sigma_{z}^{(j)} \! + \!\! \sum_k \! \omega_k b_k^{\dagger}b_k \! + \! \sum_{j=1}^2	 \! \sum_k \! \sigma_{z}^{(j)} \!\! \left(g_k b_k^{\dagger}+g_k^{*}b_k \! \right)
\end{equation}
where again we assume that the two qubits have the same couplings $g_k$ to the environment. Moving to the interaction picture and calculating the reduced dynamics of the two qubits, one obtains:
\begin{equation}
\rho_c(t)=\Phi_c(t)\circ\rho(0)
\end{equation}
where the map is
\tc{\begin{equation}
\! \Phi_{\!c}(t) \! = \! \left( \! \begin{matrix}
1 & e^{-\Gamma(t,\omega_c)}&e^{-\Gamma(t,\omega_c)}&e^{-4\Gamma(t,\omega_c)} \\ e^{-\Gamma(t,\omega_c)} & 1&1&e^{-\Gamma(t,\omega_c)}\\e^{-\Gamma(t,\omega_c)}&1&1&e^{-\Gamma(t,\omega_c)}\\e^{-4\Gamma(t,\omega_c)}&e^{-\Gamma(t,\omega_c)}&e^{-\Gamma(t,\omega_c)}&1 \end{matrix}
\right)
\end{equation}}
and $\Gamma(t,\omega_c)$ is defined in Eq. \eqref{gamtau}.
\section{Local quantum estimation theory}\label{sec:3}
Consider a family of quantum states $\rho_{\omega_c} $ depending on an unknown  parameter $\omega_{c}$. In order to infer the value of the parameter  we perform a large number of repeated measurements on the system and then process the outcomes to build an estimator $\hat{\omega}_c$ for the parameter. This procedure will inevitably associate an error to the estimator, that can be quantified through its variance $\sigma^2$.
Local quantum estimation theory (LQET) tells us which estimation strategies lead to  precise estimators, comparing the Fisher information (FI) of a certain measurement, with the quantum Fisher information (QFI).
Indeed, there is a bound to the precision of any  unbiased estimator, given by the Cram\'er-Rao inequality:
\begin{equation}
\sigma^2(\hat{\omega}_c)\ge\frac{1}{M F(\omega_c)} 
\label{crb}
\end{equation}
where $M$ is the number of repeated measurements and $F(\omega_c)$ is the Fisher
information associated to a certain measurement whose outcomes $\{x\}$ are distributed according to the conditional probability $p(x|\omega_c)$:
\begin{equation}
F(\omega_c)=\int dx \frac{1}{p(x|\omega_c)}\left(\frac{\partial p(x|\omega_c)}{\partial\omega_c}\right)^2.
\end{equation}
The CRB \eqref{crb} can  further be bounded by the quantum Cram\'er-Rao bound (QCRB)
\begin{equation}
\sigma^2(\hat{\omega}_c)\ge\frac{1}{M H(\omega_c)} 
\label{qcrb}
\end{equation}
where we introduced the QFI $H(\omega_c)$, obtained by maximizing the FI over all possible measurements \cite{parisQE}.

The explicit expression of
the QFI can be found after diagonalizing the density matrix of the system of interest $\rho_{\omega_{c}}=\sum_{n}\rho_{n}\vert\phi_{n}\rangle\langle\phi_{n}\vert$:
\begin{equation}
H\!(\omega_{c})\!=\! \sum_n \frac{(\partial_{\omega_c} \rho_n)^2}{\rho_{n}}+
2\sum_{n\ne m}\frac{(\rho_n-\rho_m)^2}{\rho_n+\rho_m}\,\left| \langle\phi_{m}\vert\partial_{\omega_c} \phi_{n}\rangle \right| ^2
\label{qfi}
\end{equation}
where $\partial_{\omega_c}$ is the derivative with respect the parameter $\omega_c$.
 The first term in Eq. \eqref{qfi} is the classical FI of the distribution $\left\{\rho_{n}\right\}$, while the second term is quantum in its nature and
vanishes when the eigenvectors of $\rho_{\omega_{c}}$ do not depend on
the parameter $\omega_{c}$.
Another figure of merit that can be addressed in order to evaluate the precision of an estimator is the signal-to-noise ratio (SNR) $r(\omega_c)=\frac{\omega_c^2}{\sigma^2(\omega_c)}$.  This quantity is always bounded \tc{from above} by the quantum signal-to-noise ratio (QSNR), defined as:
\begin{equation}
R(\omega_c)=\omega_c^2 \, H(\omega_c).
\label{qsnr}
\end{equation}
A large value of the QSNR thus means that the parameter can be estimated efficiently, with a small error.
%
\section{Cutoff frequency estimation by quantum probes}\label{sec:4}
In this section we report our results about the estimability of the  cutoff frequency of the spectral density $J(\omega)$ belonging to the Ohmic family.  This is achieved by analyzing the 
behavior of the QFI and the QSNR for fixed values of the Ohmicity parameter $s$. 
In the case of a single qubit we are able to find the optimal preparation state, which maximizes the 
QFI, and the {\it optimal} measure, such that its FI equals the QFI, i.e. $F(\omega_c)=H(\omega_c)$.
 In the case of two qubits, we compare the QFI for different initial states, i.e. product  and entangled states, in both common and independent environments (see Table \ref{tab1}). 
 Our aim is to understand whether quantum correlations can improve the estimation precision or if a  single qubit is already sufficient for efficient estimation.  Indeed we bring evidence that a simple quantum probe like a single qubit is enough to efficiently estimate the cutoff frequency of an Ohmic spectral density in a dephasing dynamics.
\subsection{Single qubit}
In this section we analyze the estimability of the cutoff frequency  of the  spectral density belonging to the  Ohmic family \eqref{Jomega} 
using a single qubit as a quantum probe. We initially prepare the qubit in a pure state  depending upon the parameter $\theta$:
\begin{equation}
\ket{\psi_0}=\cos\!\left(\frac{\theta}{2}\right)\ket{0}+\sin\!\left(\frac{\theta}{2}\right)\ket{1}.
\end{equation}
The QFI can be analytically computed according to Eq. \eqref{qfi} after diagonalizing the  density matrix for the qubit $\rho_0=\ket{\psi_0}\!\!\bra{\psi_0}$:
\begin{equation}
H(t,\omega_{c})=\frac{\sin^2\!\theta \, [\partial_{\omega_c} \Gamma(t,\omega_c)]^2 }{e^{2\Gamma(t,\omega_c)}-1}
\label{h1b}
\end{equation}
which is maximized for $\theta=\frac{\pi}{2}$ such that the optimal initial state preparation is 
$\ket{+}=\frac{1}{\sqrt{2}}(\vert0\rangle+\vert1\rangle)$,  \tc{independent on the value of $\omega_c$ and the interaction time.}
\begin{figure}
 \centering
     \includegraphics[width=1.01\columnwidth]{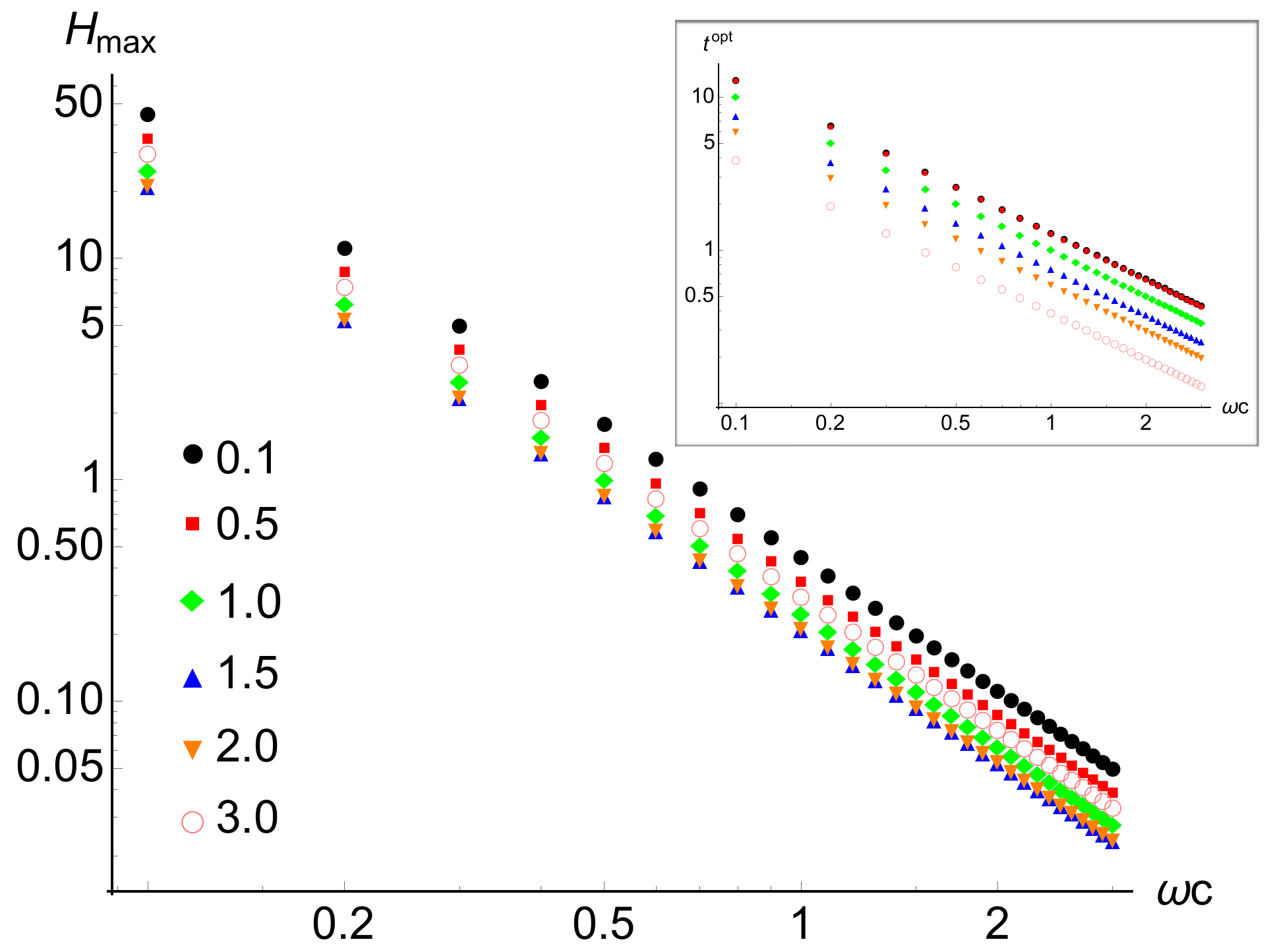}
\caption{\label{opt1}  Quantum Fisher information  $H_{\text{max}}$ and optimal time $t^{\text{opt}}$ (inset) as a function of $\omega_c$ for different values of the parameter $s$ (in the legend), in the single-qubit case. }
 \end{figure}
We recognizes that the QFI coincides with the FI
of population measurement  of the qubit  diagonalized density matrix \cite{benedetti14}:
\begin{equation}
H(t,\omega_{c})=\frac{[\partial_{\omega_c} \Gamma(t,\omega_c)]^2 }{e^{2\Gamma(t,\omega_c)}-1}.
\label{QFsingle}
\end{equation}
By substituting the the explicit form of $\Gamma(t,\omega_c)$ \eqref{gamtau} into the above equation, one gets the analytical expression  of the decoherence coefficient for fixed values of $s$.

\begin{figure}
 \centering
 \includegraphics[width=\columnwidth]{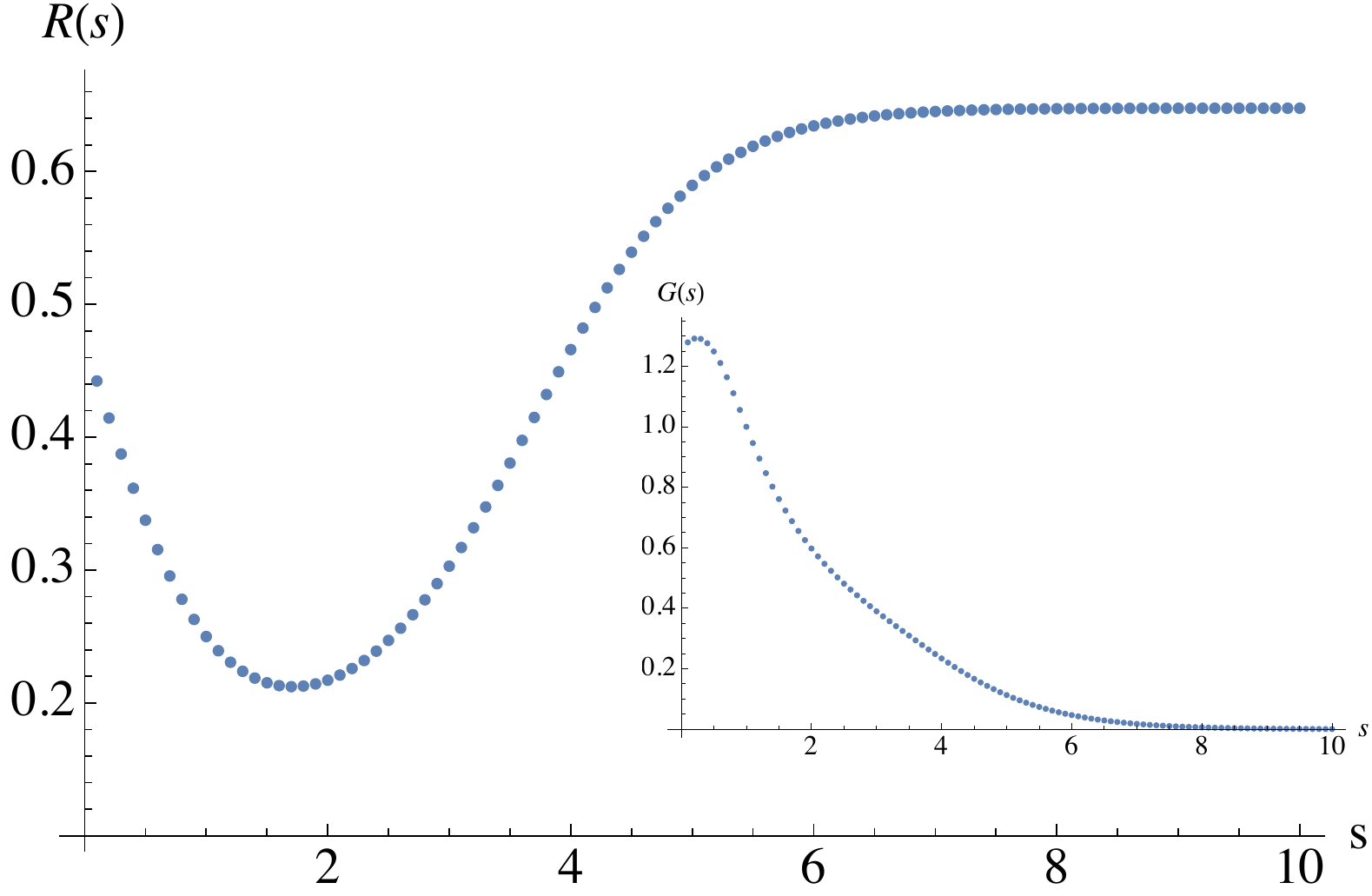}
\caption{\label{fig2} Dependency of the QSNR $R$ on the parameter $s$ for the single-qubit case. In the inset we 
report the behavior of the coefficient $G$ as a function of $s$.}
 \end{figure}

In order to optimize the inference procedure, we look for
the interaction time that maximizes the QFI as a function of $\omega_c$ and  for  a fixed value of $s$.
The maximization of the QFI over time has been performed numerically.
The optimal time $t^{\text{opt}}(s,\omega_c)$, where the quantum Fisher information has a maximum for every values of $s$, is inversely proportional to the cutoff frequency while the quantum Fisher information calculated at the optimal time is inversely proportional to the square of $\omega_c$:
\begin{equation}
t^{\text{opt}}(s,\omega_c)=\frac{G(s)}{\omega_{c}}\qquad H(t^{\text{opt}},s,\omega_c)=\frac{R(s)}{\omega_c^2},
\label{ottimi1}
\end{equation}
as shown in Fig. \ref{opt1}.
The quantity $G(s)$ does not depend on the value of the parameter to be estimated $\omega_c$, but only on the Ohmicity $s$., When we substitute the optimal time into the expression for $H$, we obtain that the optimized QFI scales with the inverse of $\omega_c^2$. This means that the QSNR $R(t^{\text{opt}},s,\omega_c)=\omega_c^2 H(t^{\text{opt}},s,\omega_c)$ is independent of the value of $\omega_c$ since it depends only on the parameter $s$.
The QSNR  has the expression:
 \begin{align}
  \!\!R(s)\! = 
  \! \left\{ 
  \begin{array}{ll}
  \!\! \frac{G^2(s)}{\big(1+G^2(s)\big)^2}
  &s\!=\!1\\
  \\
  \!\!\! 
   \frac{ 
   \coth\left[\!\!\left(1 - 
        \frac{ \cos[(s-1) \arctan G(s)]}{\big(1 +  G^2(s)\big)^{\frac{s-1}{2}}}\right) \bar{\Gamma}[s-1]\! \right]-1 }
         {2\frac{ \big(1+G^2\!(s)\big)^s}{G^{2}\!(s) \bar{\Gamma}[ s]^{2}}  \csc^2 \! [s \arctan G(s)]} 
  &s\!\neq\!1
  \end{array}
  \right. ,
  \label{r1}
  \end{align}
where $G(s)$ is the proportionality constant of the optimal time \eqref{ottimi1}. Both the $G(s)$ and $R(s)$ are reported in Fig. \ref{fig2}, which shows us that $R(s)$ has a non-monotone behavior in $s$, with a global minimum. 
The fact that $R(s)$  is independent on the value of \tc{$\omega_c$} means
that using a single qubit as a quantum probe allows  a uniform estimation of the
cutoff frequency. 
\tc{For small values of $s$ the QSNR decreases, then it reaches a minimum after which it starts increasing until it saturates to a constant value for large values of $s$.}
 %
  %
\begin{figure}[t]
 \centering
   \includegraphics[width=\columnwidth]{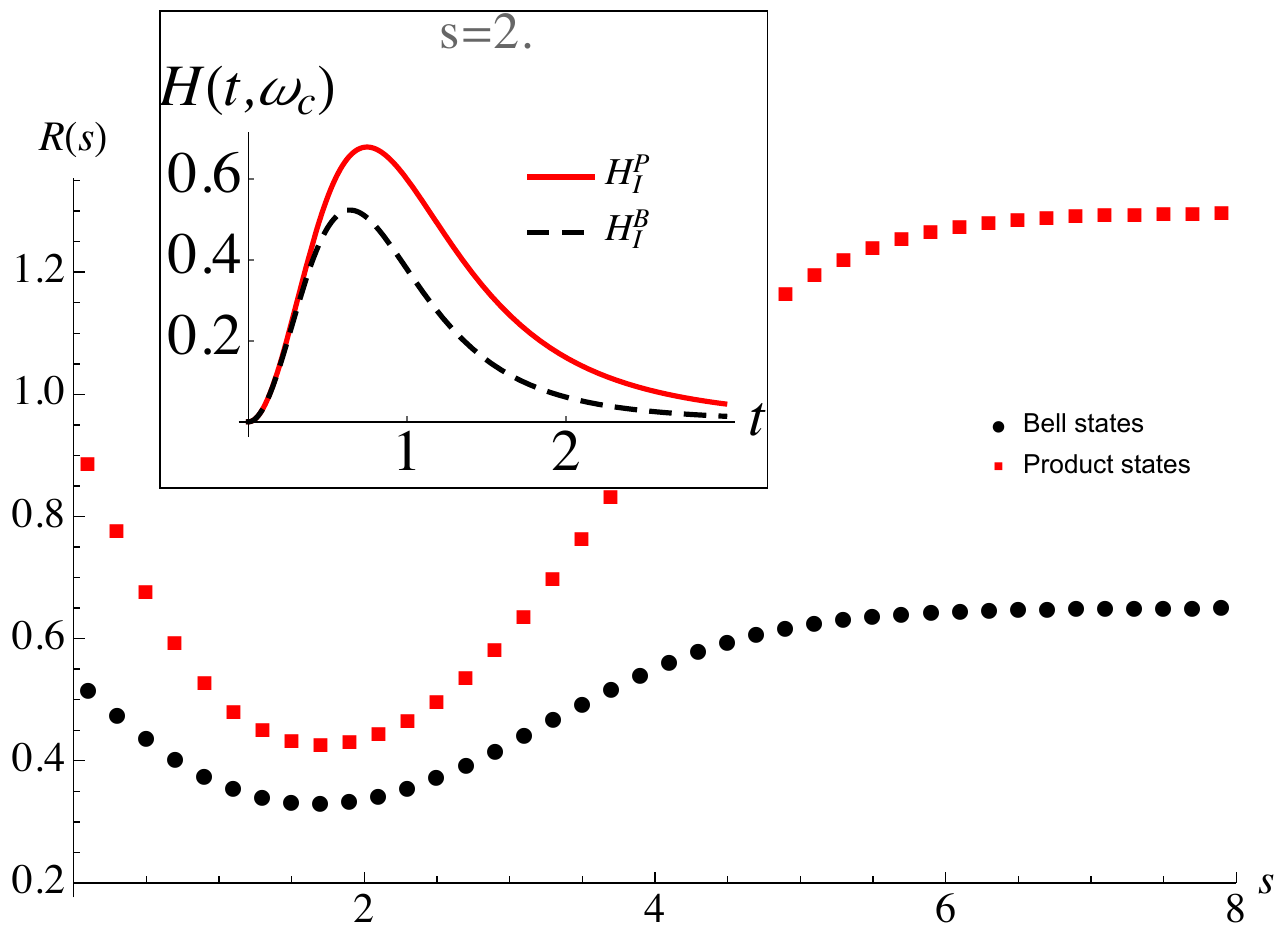}
\caption{\label{fig3} Dependency of the QSNR $R$ on the parameter $s$, in the case of two qubits interacting with identical independent baths. \tc{In the inset we compare the QFI for product (red solid line) and Bell (black dashed line) states as a function of time for $\omega_c=0.8$}.}
 \end{figure}
 \subsection{Two qubits}
 We now focus on the situation where two qubits are used as quantum probes, in order to understand whether multiple quantum probes perform better than a single qubit.
The maximization over a generic initial state of the qubits is not trivial in this case. For this reason we focus on two different state preparations, i.e. the four product states $\ket{\pm \pm}$, $\ket{\pm \mp}$ and the four Bell states $\ket{\phi^{\pm}}$ and $\ket{\psi^{\pm}}$, where $\ket{\phi^{\pm}}=\frac{1}{\sqrt{2}}(\ket{00}\pm\ket{11})$ and $ \ket{\psi^{\pm}}=\frac{1}{\sqrt{2}}(\ket{01}\pm\ket{10})$.
 Moreover, different scenarios are considered: we will start with the case where two qubits interact
 with independent local reservoirs and then we will analyze the case of two qubits in a common bath.
 \\
In the case of two qubits in independent environments (Table \ref{tab1} (a-b)), we find that all four product states lead to the same QFI, which is twice the single-qubit QFI $H(t,\omega_c)$ of Eq. \eqref{h1b}, thus confirming the additivity of the quantum Fisher information:
\begin{align}
H_I^P(t,\omega_c)=2 H(t,\omega_c).
\end{align}
Also in the case where the two qubits are initially entangled, the  QFI is the same for all four Bell states,
and it reads:
\begin{align}
H_I^B(t,\omega_c)=4\frac{[\partial_{\omega_c}\Gamma(t,\omega_c)]^2}{e^{4\Gamma(t,\omega_c)}-1}.
\end{align}
After maximizing both $H_I^P(t,\omega_c)$ and $H_I^B(t,\omega_c)$ over time, we find the same dependency as in the case of the single qubit: the optimal time is inversely proportional to the cutoff frequency and the maximized QFI scales as $\omega_c^{-2}$, as reported in Eq. \eqref{ottimi1}.
 \tc{The optimal time for product states is always  larger than $t^{\text{opt}}$ for Bell states  but 
if we fix a target precision much smaller than the QCRB, product  and Bell states can achieve it at the same time (
shown in the inset of Fig. \ref{fig3}), while an intermediate precision will be obtained faster by employing product states.
Indeed, for small times $t\ll1$, we can expand in series to third order the QFI in both cases:\begin{align}
&H_I^P\!(t,\omega_c)\!=\!2\bar{\Gamma}[1\!+\!s]t^2\!-\!\frac{\omega_c^2}{2}\!\left(2\bar{\Gamma}[1\!+\!s]^2\!+\bar{\Gamma}[3\!+\!s]\right)\!t^4\\
&H_I^B(t,\omega_c)\!=\!2\bar{\Gamma}[1\!+\!s]t^2\!-\!\frac{\omega_c^2}{2}\!\left(4\bar{\Gamma}[1\!+\!s]^2\!+\bar{\Gamma}[3\!+\!s]\right)\!t^4
\end{align}
and we see that up to second order the two expansions coincides.\\
 The QFI at its optimal time is always higher for initial product states than for Bell states for a fixed value of $\omega_c$ and since it
 is proportional to $\omega_c^{-2}$,
 }
 it follows that the QSNR is constant and depends only on $s$. \tc{Since the we are interested in the maximum precision allowed by quantum mechanics, we consider as a figure of merit to the goodness of the inferring procedure the QSNR evaluated at the optimal time.}
 In Fig. \ref{fig3} we compare the behavior
 of $R(t^{\text{opt}},s,\omega_c)$ for initial product and Bell states.
\begin{figure}[t]
 \centering
 \includegraphics[width=1\columnwidth]{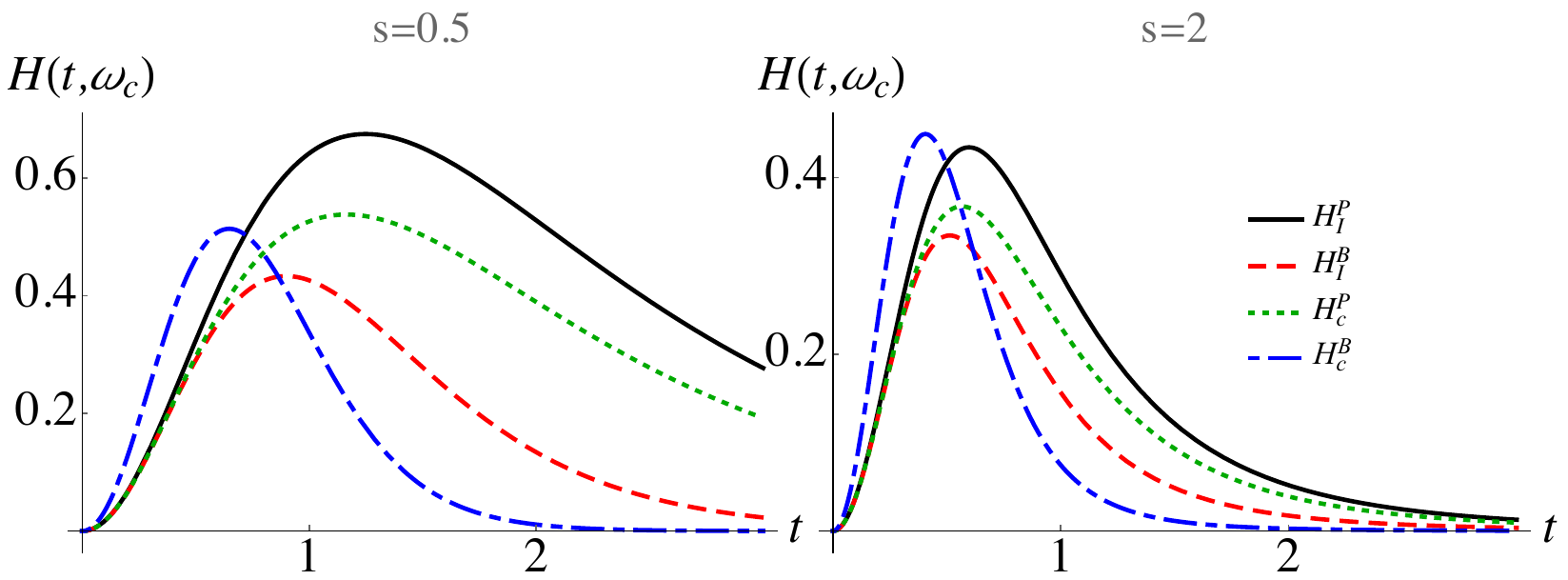}\\
 \includegraphics[width=.98\columnwidth]{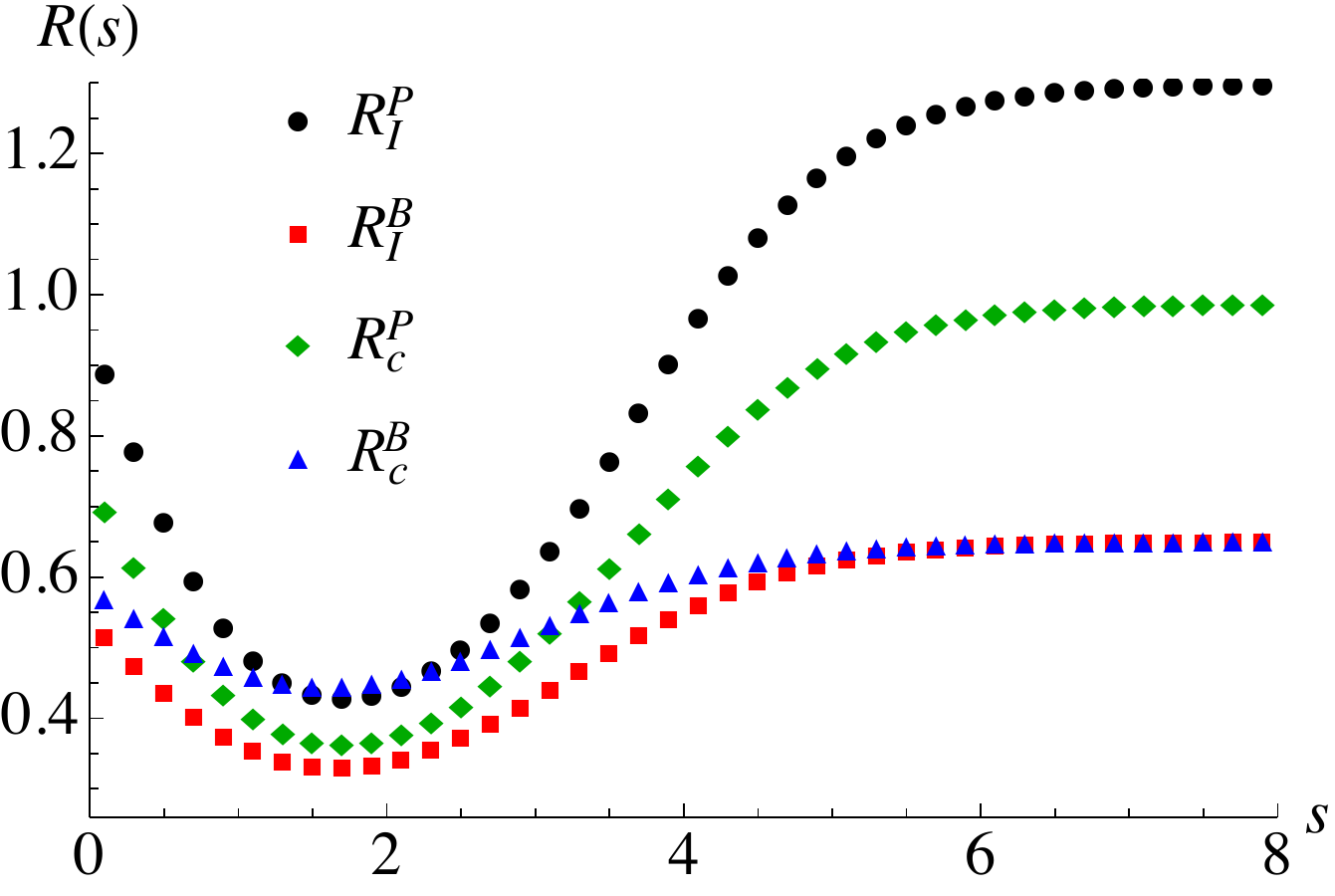}
 \caption{\tc{(Top) Fisher information as a function of time for two fixed values of the parameter $s$ in the cases of two qubits in independent environments in product (black solid line) or Bell (red dashed line) state  and in a common environment prepared in a product (green dotted line) and Bell (blue dot-dashed line) state for $\omega_c=1$. (Bottom)
 } Comparison between the QSNR as a function of the Ohmicity  $s$, obtained from the optimized quantum Fisher information, for four different initial conditions of the qubit:  two qubits initially in a separable (black dots) or an entangled (red squares) state in  independent reservoirs and two qubits in a common environment initialized in a product (green diamonds) or an  entangled (blue triangles) state.}
  \label{fig4}
 \end{figure}
As it is apparent from the plot, the quantum correlations \tc{of Bell states} do not help in estimating the unknown parameter.
Indeed product states allow us to obtain a larger QSNR for a fixed values of the Ohmicity $s$, i.e. a 
more precise inference of $\omega_c$.
  \\
 We now consider the case where the two qubits interact with the same environment, as shown in the table \ref{tab1} (c-d).
 All four product states will give the same QFI:
\tc{  \begin{align}
 H_c^P(t,\omega_c)&=[\partial_{\omega_c}\Gamma(t,\omega_c)]^2\,\times\nonumber\\
&
 \frac{8\left\{1+e^{4\Gamma(t,\omega_c)}\left[1+\sinh(2\Gamma(t,\omega_c))\right]\right\}}
 {3 e^{8\Gamma(t,\omega_c)}-2e^{4\Gamma(t,\omega_c)}-1}
  \end{align}}
  while for Bell states, only the $\ket{\psi^{\pm}}$ give a significant contribution, with a QFI equal to:
\tc{\begin{align}
 H_c^B(t,\omega_c)= \frac{16\, \left[\partial_{\omega_c}\Gamma(t,\omega_c)\right]^2}{e^{8\Gamma(t,\omega_c)}-1},
 \end{align}}
As before, \tc{we are interested in the the optimized QFI: we find that it is inversely proportional to $\omega_c^2$, such that the QSNR is constant 
for a fixed value of $s$.}
\tc{Our previous result hold true for most of the values of the parameter $s$, i.e. product states in independent baths yield the higher value of the QSNR compared to the other scenarios. However, there exists a range of values of the Ohmicity parameter for which the $R(s)$ is larger if we employ the Bell states in a common-bath scheme. This is shown in Fig. \ref{fig4}, where we compare the behavior of the QFI for two different values of $s$ as a function of time in the four estimation schemes considered in this paper (top plots) and the QSNR as a function of $s$ (bottom plot). In particular, we emphasize the fact that there exist values of $s$, such as $s=2$ in our example, where the estimation of the cutoff frequency  is improved if we employ  a common-bath scheme  with two qubit in a Bell state. Indeed,  we see that $H^B_c(t^{\text{opt}},\omega_c)$ is larger than $H^P_I(t^{\text{opt}},\omega_c)$.
 It is also worth noticing that in this case the optimal time for $H^B_c$ is shorter than that of independent probes.   }

\begin{widetext}

\centering
 \begin{table}[!h]
 \centering
  \begin{tabular}{  |c|c|c|c|}
    \hline
  \rule{0pt}{12pt}  (a) & (b) & (c)&(d) \\[3pt]
  \hline
   \begin{minipage}{.2\textwidth} \centering
      \includegraphics[width=.9\textwidth]{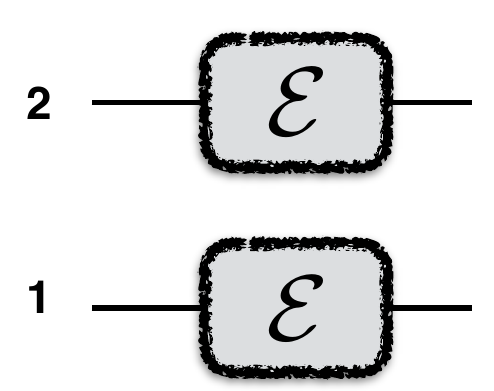}
    \end{minipage}&\begin{minipage}{.2\textwidth}\centering
      \includegraphics[width=0.88\columnwidth]{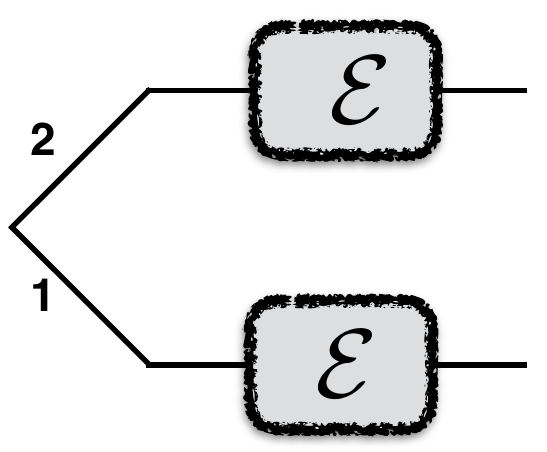}
    \end{minipage}  & \begin{minipage}{.2\textwidth}\centering
      \includegraphics[width=0.84\columnwidth]{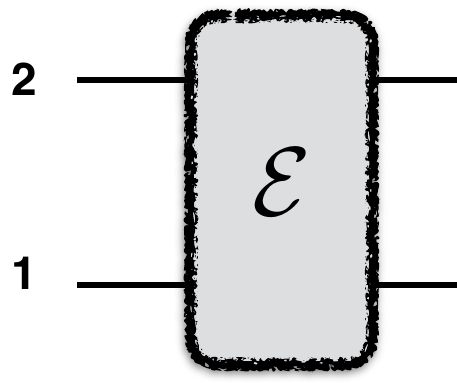}
    \end{minipage} &  \begin{minipage}{.2\textwidth}\centering
      \includegraphics[width=0.9\columnwidth]{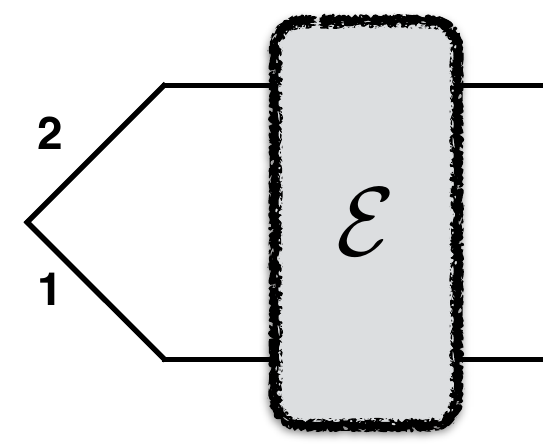}
    \end{minipage}   \\[33pt]
    \hline 
  \rule{0pt}{23pt} 
   $H_I^P=2 \, \dfrac{[\partial_{\omega_c}\Gamma(t,\omega_c)]^2}{e^{2\Gamma(t,\omega_c)}-1}$ & 
  $
  H_I^B=4\dfrac{\left[\partial_{\omega_c}\Gamma(t,\omega_c)\right]^2}{e^{4\Gamma(t,\omega_c)}-1}$ 
  &
  $\tc{  
 H_c^P\!\!= \!\!
 \frac{8\left\{\!1+e^{4\Gamma\!(t,\omega_c)}\left[1+\sinh(2\Gamma\!(t,\omega_c))\right]\!\right\}[\partial_{\omega_c}\!\Gamma(t,\omega_c)]^2}
 {3 e^{8\Gamma(t,\omega_c)}-2e^{4\Gamma(t,\omega_c)}-1}
 } $ 
  &
   $\tc{H_c^B=  \frac{16\, \left[\partial_{\omega_c}\Gamma(t,\omega_c)\right]^2}{e^{8\Gamma(t,\omega_c)}-1}}$
  \\[10pt]
  
  \hline
  
  \end{tabular}
  \caption{
Summary of results for two-qubit quantum probes.
We compare four different estimations schemes for the cutoff frequency of the spectral density  $J(\omega)$ in Eq. \eqref{Jomega} using two qubits as quantum probes: (a)  qubits prepared in a separable state interacting with independent and identical reservoirs,(b) qubits prepared in a Bell state interacting with independent and identical reservoirs,(c) qubits prepared in a separable state coupled to a common bath,  (d) qubits prepared in a Bell state coupled to 
 a common bath. We also report the expressions for  their respective QFI as a 
 function of the decoherence factor $\Gamma$.
   }\label{tab1}
\end{table}

\end{widetext}
Since employing two non-interacting qubits that are coupled to independent identical reservoirs  initialized in a separable state is the same as repeating twice the single-qubit procedure described in section \ref{sec:2}, it follows that 
\tc{using a single qubit as a probe is sufficient to optimally 
estimate the cutoff frequency of an ohmic spectral density for most values of $s$.
This is due to the fact that, for those values of $s$, using multiple qubits in a Bell state, in common or independent reservoirs, does not lead to improvements in the estimation procedure. Common values for $s$  are $s=\frac12$, $s=1$ and $s=3$ \cite{paavola}, and they fall into this case, where a single qubit is the optimal probe.
This is a relevant conclusion, that tells us that the simplest quantum probe, 
a qubit, is sufficient to estimate the spectral parameter of the environment. However, we also found that there is a small range of  the Ohmicity parameter where it is better to use two qubits prepared in a Bell state interacting with the same quantum bath, in order to obtain a larger estimation precision.}
\\
\tc{
In order to deepen our analysis to include states with a different amount of entanglement, we analyze 
the performances of Werner states $\rho_W$ as quantum probes, 
where $\rho_W=p\ketbra{\phi^B}{\phi^B}+(1-p)\mathbb{I}/4$ with  $I$  the identity 
matrix and $\ket{\phi^B}$ one of the four Bell states. The parameter $p$ is related to the purity $P$
of the state through the relation $P=(1+3p^2)/4$ and the associated entanglement $E$ is nonzero only for 
$1/3<p<1$ and is $E=(3p-1)/2$. The QFI for two qubits initialized in a Werner states and interacting with separate bath or a common environment takes the expression:
\begin{align}
H_I^W(t,\omega_c,p)&=\frac{8p^2(1+p)\,[\partial_{\omega_c}\Gamma(t,\omega_c)]^2}{(1+p)^2e^{4\Gamma(t,\omega_c)}-4p^2}\\
H_c^W(t,\omega_c,p)&=\frac{32p^2(1+p)\,[\partial_{\omega_c}\Gamma(t,\omega_c)]^2}{(1+p)^2e^{8\Gamma(t,\omega_c)}-4p^2}.
\end{align}
}
\begin{figure}[t]
 \centering
  \includegraphics[width=1.05\columnwidth]{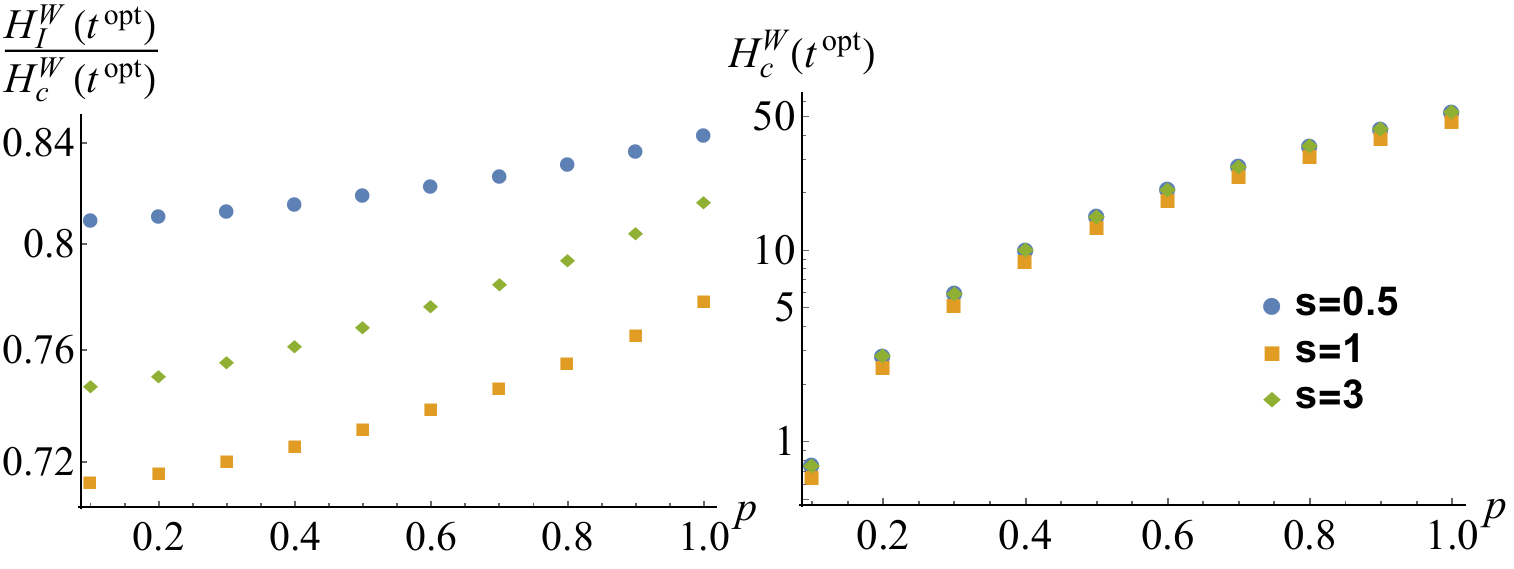}
 \caption{\tc{(Left) Ratio of the optimized QFI for Werner states in independent and common baths, as a function of the parameter $p$, for three different values of $s$. (Right) Behavior of
 the optimal QFI for two qubit in a Werner state interacting in with a common environment for $\omega_c=0.8$. }}
  \label{fig5}
  \noindent
 \end{figure}
\tc{
Figure \ref{fig5} shows the ratio $\frac{H^W_I(t^{\text{opt}})}{H^W_c(t^{\text{opt}})}=\frac{R_I^W}{R_c^W}$ as a function of the parameter $p$ and for three different values of $s$.  We first notice that, since the ratio is smaller than 1, Werner state  perform better in a common bath than in independent environments. Since Bell states ($p=1$) permit one to achieve the largest precision, we can exclude the use of Werner states as optimal quantum probes and no improvement is gained in their use.
\\
\\
At last, we can ask ourselves what happens if we use $N$ qubit as a probe. 
As in the case of two qubit, the generalization to $N$ qubits cannot be done analytically, except for few selected cases.
Here we extend the analysis of the QFI for Greenberger-Horne-Zeilinger (GHZ) states $\ket{\psi_{GHz}}=\frac{1}{\sqrt{2}}(\ket{000\dots}+\ket{111\dots})$ in independent and common baths, postponing a more complete discussion for future works. The QFI for the GHZ states reads:
\begin{align}
H_I^{\text{GHZ}}(t,\omega_c,N)&=\frac{N^2 [\partial_{\omega_c}\Gamma(t,\omega_c)]^2}{e^{2 \,N\, \Gamma(t,\omega_c)}-1}\\
H_c^{\text{GHZ}}(t,\omega_c,N)&=\frac{N^4 [\partial_{\omega_c}\Gamma(t,\omega_c)]^2}{e^{2 \,N^2\, \Gamma(t,\omega_c)}-1}.
\end{align}
The maximum of the QFI increases with the number of qubits $N$ and $H_c^{\text{GHZ}}(t,\omega_c,N)$ is larger than $H_I^{\text{GHZ}}(t,\omega_c,N)$ for fixed values of the parameters. However these QFI remain smaller than the quantum Fisher information obtained using $N$ independent qubits as probes.
}
%
\section{Conclusions}\label{sec:5}
In this paper we have addressed the estimation of the
cutoff frequency of an Ohmic reservoir
using single-qubit and two-qubit quantum probes. The reservoir is made of an ensemble 
of non-interacting bosonic modes and the interaction between system and environment 
generates a dephasing map. We have evaluated the quantum Fisher
information for different initial states of the probes, showing that for a single-qubit probe,  
the optimal  state preparation is  the superposition  $\ket{+}$, and that the optimal 
interacting time is inversely proportional to the cutoff frequency itself $\omega_c$, such
that the maximized QSNR is independent of the value of the cutoff frequency for any 
fixed value of the Ohmicity parameter $s$. 
\par
In order to understand if multiqubit quantum probes perform better than a single-qubit one, 
we also employed two non-interacting qubits  to infer the value of $\omega_c$. 
\tc{Clearly, we can only compare specific initial states for the two-qubit case since we cannot provide the analytic expression for the two-qubit QFI for a generic initial state. For this reason we focussed only on initial product and Bell states.}
In particular, 
we compare the precision, i.e. the QFI, obtained from four different scenarios, reported in Table
\ref{tab1}.
\tc{
We showed that also in these cases the QSNR does not depend on the value of $\omega_c$ and that for most values of $s$, including the most common cases $s=0.5,\,1,\,3$ \cite{paavola}, product states perform better than Bell states in estimating the cutoff frequency. This means that a single qubit is  already optimal to  infer  the value of the cutoff frequency.  However we found that there exists a small range of parameter  approximately between $1.35<s<2.3$ where using a common environment scheme with two qubits initialized in a Bell state allows one to achieve a better estimation precision.
   } 
\par
Our work paves the way for future developments, which include the 
estimation of the spectral parameters for an Ohmic reservoir 
at non-zero temperature and the study of system-bath couplings
with different spectra.
\acknowledgements
This work has been supported by EU through the collaborative H2020 project QuProCS 
(Grant Agreement 641277). FSS thanks the QTLab group
for the kind hospitality during her stay in Milan.

\end{document}